\documentclass[
    ,final            
  ]
  {aipproc}

\layoutstyle{6x9}

\newcommand{\nm}{\ensuremath \mbox{nm}}
\newcommand{\um}{\ensuremath \mu\mbox{m}}
\newcommand{\mm}{\ensuremath \mbox{mm}}
\newcommand{\m}{\ensuremath \mbox{m}}
\newcommand{\mHz}{\ensuremath \mbox{mHz}}
\newcommand{\Nm}{\ensuremath \mbox{Nm}}

\title[TB for LISA]{High Sensitivity Torsion Balance Tests for LISA Proof Mass Modeling}
\author{S.~Schlamminger}     
{address={Department of Physics, Center for Experimental Nuclear
    Physics and Astrophysics, Box 354290, University of Washington,
    Seattle, WA, 98195-4290}}
\author{C.~A.~Hagedorn}         
{address={Department of Physics, Center for Experimental Nuclear
    Physics and Astrophysics, Box 354290, University of Washington,
    Seattle, WA, 98195-4290}}
\author{M.~G.~Famulare}        
{address={Department of Physics, Center for Experimental Nuclear
    Physics and Astrophysics, Box 354290, University of Washington,
    Seattle, WA, 98195-4290}}
\author{S.~E.~Pollack}          
{address={Department of Physics, Center for Experimental Nuclear
    Physics and Astrophysics, Box 354290, University of Washington,
    Seattle, WA, 98195-4290}}
\author{J.~H.~Gundlach}         
{address={Department of Physics, Center for Experimental Nuclear
    Physics and Astrophysics, Box 354290, University of Washington,
    Seattle, WA, 98195-4290}}


\begin{abstract}
We have built a highly sensitive torsion balance to investigate small forces
between closely spaced gold coated surfaces. Such forces will occur between
the LISA proof mass and its housing. These forces are not 
well understood and experimental investigations are imperative. 
We describe our torsion balance and present the noise of the system. A
significant contribution to the LISA noise budget at low frequencies
is the fluctuation in the surface potential difference between the
proof mass and its housing. We present first results of these
measurements with our apparatus.  
\end{abstract}

\begin{document}
\maketitle

\section{Introduction}

One of the leading contributions to the LISA acceleration noise budget
at frequencies below 1~mHz 
is electrostatic force noise acting on the proof
mass~\cite{PPA,DRS_ITAT,FTR}.  Very 
little is known about spurious electrostatic forces that arise
between two conducting surfaces.  

We have built a highly sensitive torsion balance apparatus to measure
small forces which may act between closely spaced surfaces.  
We introduce our instrument, its performance and an investigations of
these electrostatic forces.

\section{Experimental setup}

Our instrument consists of a vertical planar metal surface parallel to
which we suspend a thin plate from a thin wire. The latter plate is
designed to emulate LISA's proof mass housing surface and the opposing
metal surface  has properties which are similar to those of the proof
mass. The suspended thin plate forms a torsion pendulum which is
exquisitely sensitive to small forces.  

The main body of the pendulum plate is a rectangular section, cut from
a $0.43\;\mm$ thick silicon wafer, which is $l=114.3\;\mm$ long and 
$h=38.1\;\mm$ high. At the center the plate  has another smaller
rectangular 
section, $25.4\;\mm$ long and $34.9\;\mm$ high, so that the pendulum has the
shape of an upside-down T.   At the top of the pendulum plate an
aluminum compensator bar is attached perpendicular to the plate. This bar was
designed to minimize the gravitational quadrupole moment of the
pendulum in order to
reduce the sensitivity to mass movements in its vicinity. A tungsten
torsion fiber is attached to the compensator 
bar. The fiber is $0.534\;\m$ long and has a diameter of $13\;\um$
which provides a restoring torque of $\kappa=774\;\mbox{pNm}/\mbox{rad}$.

The entire pendulum is gold coated. The silicon
plate was coated with an adhesion layer of $20\;\nm$ TiW shortly after
it was dipped into hydrofluoric acid to remove the natural oxide
layer. A $225\;\nm$ thick gold coating gold was sputtered onto  the
adhesion layer.

The metal plate parallel to the pendulum plate is mounted on a
translation stage to vary its separation from the pendulum between $0\;\mm$
to $10\;\mm$. This plate is machined from OFHC copper. It is larger than
the main plate of the pendulum: $50.8\;\mm$ high and   $127.5\;\mm$ wide. 
The copper plate is split into two halves. The halves are separated by
a 0.5~mm gap to isolate them electrically and thermally from one another. To
each copper plate an electrostatic potential can be independently
applied. Each half of the copper plate  has a $600\;\mbox{mW}$ resistive heater
built-in and is equipped with  a temperature sensor. 

The two half of the copper plates have been coated by a sputtering
process similar to the one used for the pendulum. In order to prevent
the gold from diffusing into the copper we used TiW as a diffusion barrier.

\begin{figure}[h!]
\includegraphics[scale=0.45]{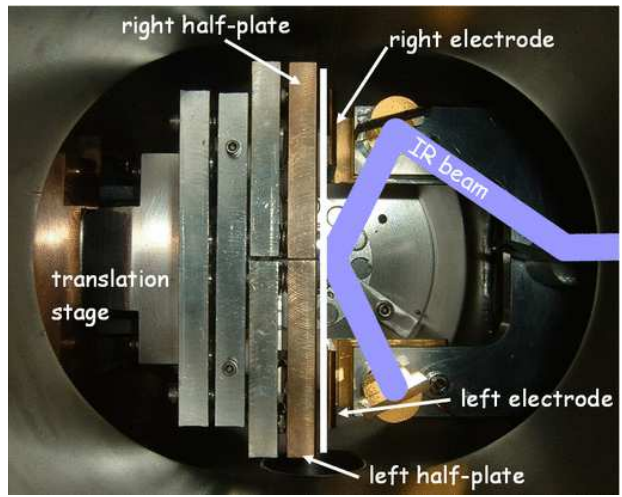}
\caption{
  Annotated photograph of the plate and  electrode layout. 
    The pendulum has been removed to take this
  picture. The pendulum would be suspended in the space between the
  copper plate and the electrodes, indicated by the white
  line. For this picture, however, the plate was moved a few mm beyond the
  pendulum's position. The light gray trace schematically shows the beam
  path of the autocollimator located to the right outside the vacuum
  vessel.
}
\label{fig:topview}
\end{figure}

In order to control the torsional motion of the pendulum, two electrodes (right and left) were
installed near the back of the pendulum plate (see
Figure~\ref{fig:topview} for a top view of the geometry). These electrodes
are small (25~mm by 40~mm high) and far enough away ($\approx
6$~mm) that they do not cause any excess noise. On each electrode a 
voltage can be independently applied. The pendulum itself is grounded
through the torsion fiber which has a resistance of $270\;\Omega$. 
 
The rotational angle of the pendulum is measured with an
autocollimator. The infra-red autocollimator beam passes through a
glass window into the vacuum chamber and is reflected off the pendulum
twice before it is returned to the autocollimator. The return beam is
focused onto a position-sensitive photo-detector.  
  
The pendulum and the plate are in a stainless steel vacuum
chamber that is maintained with a turbo-molecular pump at
$\approx 10^{-5}$~Pa. Systematic studies of pressure-related effects were
done by changing the rotation rate of the turbo pump~\cite{Scott06}.  

The actuator for translating the copper plate is located
outside the vacuum chamber and has a resolution of $0.001\;\mm$. In order to
measure the absolute distance between the pendulum and the plate three
different methods were used: (A) the plate was moved near the pendulum and
the clockwise and counterclockwise deflection angles at which the
pendulum touched the copper plate were measured. For 
a given distance the maximum deflection angle is given by half the
length of the pendulum divided by the separation. (B) a window
on the bottom of the vacuum chamber accommodates imaging of the pendulum with
a CCD camera. The separation can be inferred from the 
video image. (C) we  measured the capacitance between the
pendulum and the copper plate as a function of distance. By fitting the function $C(x)=A/ (x-x_o)+C_p$ we fit for the distance $x_o$. The parasitic
capacitance $C_p$ can be established by moving the plate far away from
the pendulum.  

The apparatus rests on a thick copper plate and is located inside
two layers of styrofoam. The instrument is located on a deep
foundation in a thermally stabilized room. 
 
\section{Measurements}

In order to measure torques on the pendulum we employ two different
methods. 

In the so-called free-run mode, the pendulum is allowed to move
freely and the torsional angle as a function of time is
recorded. This mode is used to assess the baseline noise of
the system, i.e. at plate-pendulum separations >$1\;\mm$. At smaller
separations the ambient torques on the pendulum were too large for
this mode of operation. The second mode has the pendulum held in
position using the control electrodes. A digital 
PID control loop applies voltages to either one or the other of
the control electrodes so that the autocollimator reading remains
essentially unchanged. The torque on the pendulum is proportional to
the control electrode voltage squared. The voltage-to-torque
calibration depends on the electrode and pendulum geometry. Therefore,
we developed a technique which allows us to continuously 
monitor the instrument's calibration.

We rotate two $\approx 2.0$~kg brass cylinders, around the
pendulum outside the vacuum chamber, at constant frequency ($2.48\;\mHz$). 
The gravitational
coupling between the pendulum and these masses produces small torques at twice and four times the rotation frequency. Since these
calibration signals are at fixed frequencies, they do not affect the
science data. To
absolutely calibrate the electrostatic feedback torque we compare this
gravitational torque signal to the same torque measured in the a free
run method. The signal at four times per revolution was also
calculated using the pendulum mass distribution.   

There are two significant advantages to running the system in the feedback
mode: (1) the dynamic torque range of the system is expanded by the gain
of the feedback loop ($\approx 1000$) to $\approx 5\times 10^{-10}\;\Nm$,
and (2) the system 
recovers quickly from rare but large disturbances to the pendulum, such
as earthquakes, fiber-quakes, or spontaneous pressure spikes.

\subsection{Torque sensitivity of our torsion balance}

The noise performance of our torsion balance is best demonstrated
by the power spectral amplitude of the torque noise when the
pendulum is operated in the free running mode and with a large
plate-pendulum separation. The torque spectrum is generated from the
angular deflection spectrum by dividing by the pendulum's response
function given by 
\begin{equation}
r(f) = \frac{1/\kappa}{(1-f^2/f_o^2)+i/Q},
\end{equation}
where $f_o$ is the natural frequency and $Q$ is the oscillator's quality
factor $Q\approx 3000$.

\begin{figure}[h!]
\includegraphics[scale=1.0]{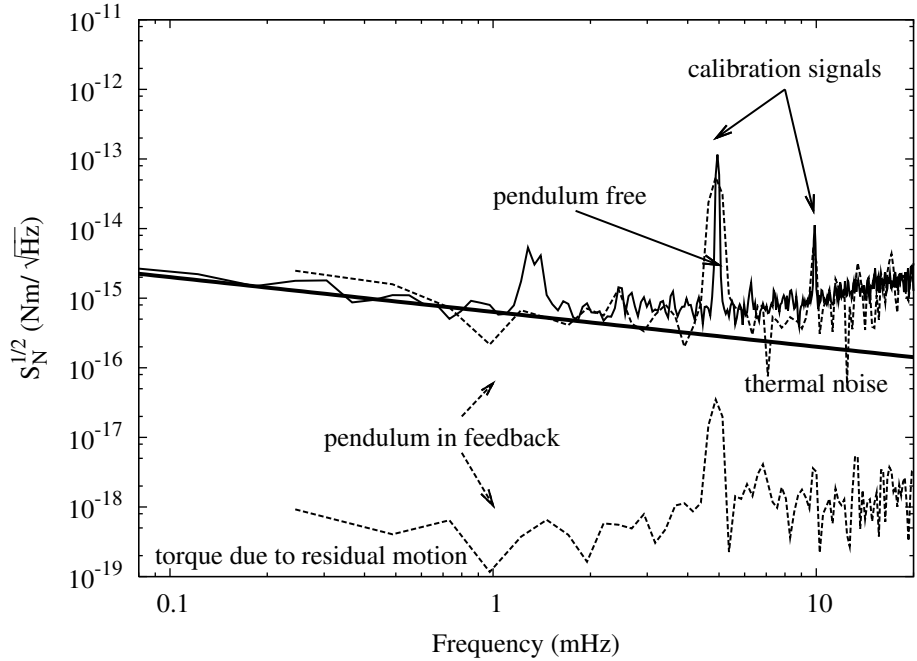}
\caption{Performance of our torsion balance. The thick line presents the
theoretical limit given by the thermal noise. The solid line is the
measured power spectral amplitude of the torque in the free running
mode. The dashed lines represent the torque noise for the system in the
feedback mode. The upper dashed trace is the torque produced by the control
electrodes and the lower trace is the torque stored in the fiber by
the residual deflection of the pendulum. This data was taken at a
plate-pendulum separation of $5\;\mm$.}
\label{fig:Torquenoise}
\end{figure}

Figure~\ref{fig:Torquenoise} shows the power spectral amplitude of the
torque as a function of frequency for a free running pendulum and for
the pendulum in feedback. For the pendulum in feedback two traces are
shown. The upper curve is the torque noise measured in feedback
and the lower curve is the remaining torque in the
fiber. At low frequencies, both measurement techniques reach
the thermal limit of a torsion balance, given by
\begin{equation}
S_N^{1/2}(f) = \sqrt{\frac{4k_BT\kappa}{2\pi Qf}},
\end{equation}
with the Boltzmann constant $k_B$ and the room temperature
$T$~\cite{Saulson90}. At higher  frequencies ($> 5\;\mHz$)  the
measured torque noise is in excess of the  
thermal noise due to  autocollimator noise, typically
$130\;\mbox{nrad}/\sqrt{\mbox{Hz}}$ at $6\;\mHz$. The gravitational
calibration signals can be 
seen at $4.95\;\mHz$ ($Q_{22}$) and $9.9\;\mHz$ ($Q_{44}$).

\begin{figure}[h!]
\includegraphics[scale=1.0]{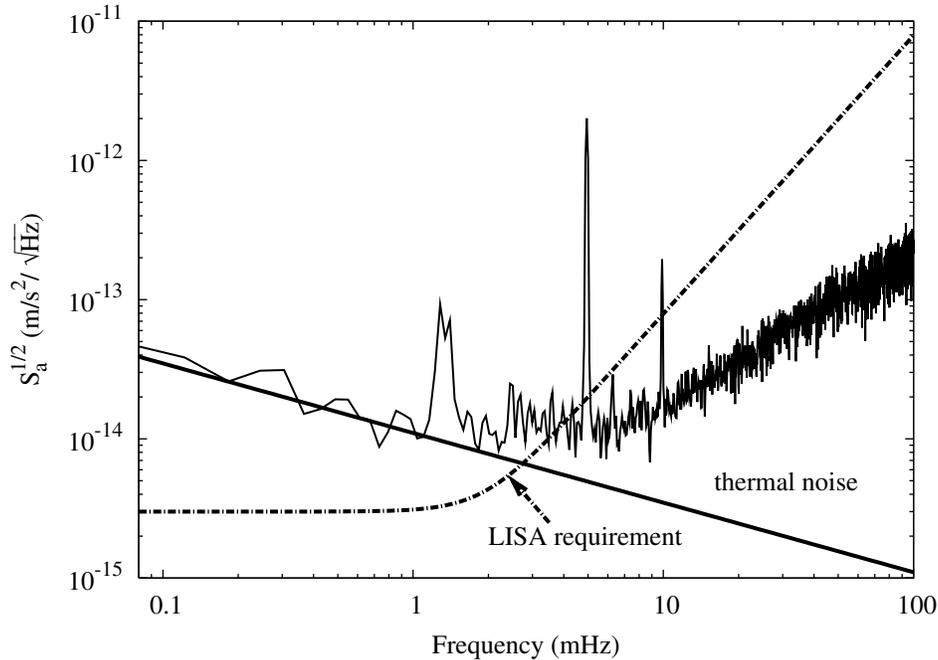}
\caption{LISA-equivalent acceleration noise as a function of
  frequency. The thick solid line is the thermal limit for our torsion
  balance. The two sharp peaks correspond to our calibration signals.}
\label{fig:ANoise}
\end{figure}

For purposes of comparison, we convert our measured torque noise
spectra into acceleration noise spectra, which can be compared with LISA
requirements. To convert torques into LISA-equivalent accelerations we
use the expression 
\begin{equation}
a = \frac{ 4 \eta}{M}\frac{A}{l^2h} N,
\end{equation}
in which $M=1.96\;\mbox{kg}$ is the mass of the LISA test mass and $A=21.16\;\mbox{cm}^2$ is the
surface area of one side of the cube, $N$ is the torque, and $\eta$ is
a geometrical integration constant, for which a conservative value is
two. 
Figure~\ref{fig:ANoise} shows the LISA-equivalent acceleration
noise. At $1\;\mHz$ the acceleration noise of 
our apparatus is one order of magnitude above the LISA
requirement. The torsion balance has its best performance at around
$10\;\mHz$, where the acceleration sensitivity is slightly better than the 
requirement.

\subsection{Surface Potential}

\begin{figure}[h!]
\includegraphics[scale=1.0]{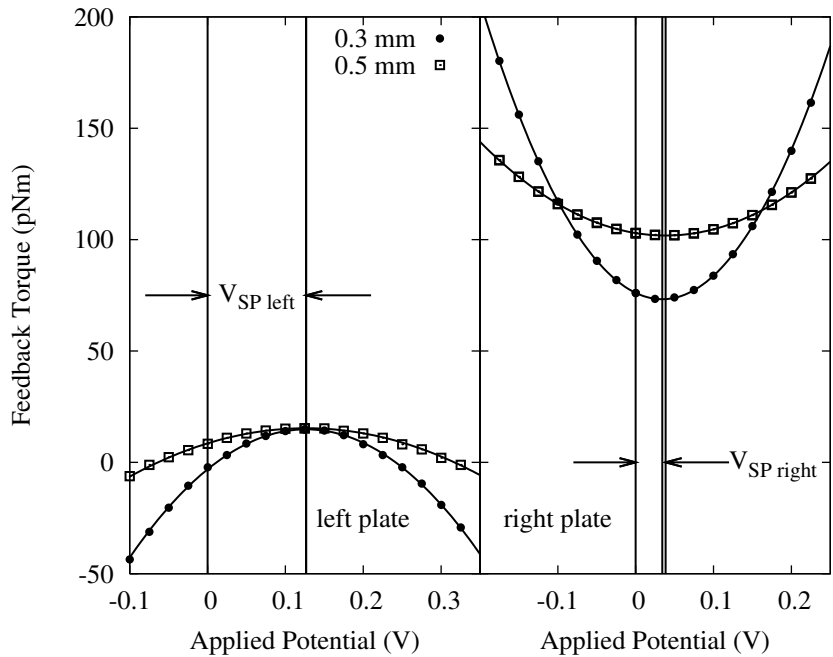}
\caption{The measured torque as a function of potential applied to each
  half plate. The potential at the minimum of each parabola gives the surface potential.}
\label{fig:parbo}
\end{figure}

Fluctuations in the electrical surface potential differences
(``surface potentials'')  are one of the dominant
noise sources at frequencies below 1~mHz. We can measure the surface  
potential between the pendulum and each half of the copper plates by using
the pendulum in feedback. 
The torque produced by a voltage applied to the copper plate is given
by $c(V-V_{SP})^2$, where $V_{SP}$ is the surface 
potential between the pendulum and the plate and $c$ is a calibration
constant, which depends on the plate-pendulum
separation. Figure~\ref{fig:parbo} shows a measurement of this 
relationship.

We found that the surface potential is 0.13~V for the left plate and
0.03~V for the right plate. The measured values for the surface
potentials are independent of the
plate-pendulum separation. However, the
surface potential was found to vary  by $\pm5\;\mbox{mV}$ in a period of 150
days. After the investigation of thermal
effects~\cite{Scott06}, the surface potential changed
drastically by $50\;\mbox{mV}$. Most of this change occurred after the first
heating cycle of the copper plate.

\begin{figure}[h!]
\includegraphics[scale=1.0]{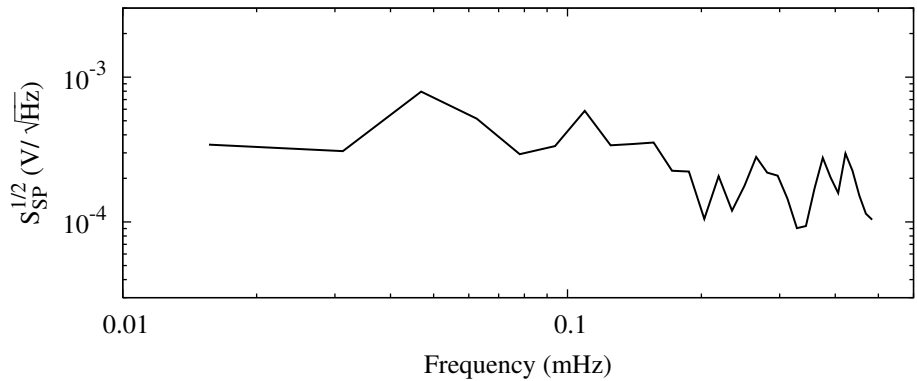} 
\caption{
A typical measurement of the power spectral amplitude of the surface
potential. This plot was generated from a 24~hour long run at a plate-pendulum separation of $0.3\;\mm$.}
\label{fig:cpfft}
\end{figure}

In order to investigate the temporal behavior of the surface potential
fluctuations we alternate between two voltages $V_1$ and $V_2$.
$V_1$ and $V_2$ were applied for 500~s each and were chosen
symmetrically around $V_{SP}$, so that 
the torque on the pendulum remains essentially  constant and transients in
the torques are minimized.  
>From small differences in the measured torque, the voltage difference
 $V_2-V_1$, and the 
 calibration constant we calculate the surface potential. 
 The power spectral amplitude of such
 a measurement is shown in Figure~\ref{fig:cpfft}. The spectrum is
 white at a level of $\approx 200\;\mu\mbox{V}/\sqrt{\mbox{Hz}}$ for
 frequencies below  $0.1\;\mHz$ and therefore most likely
 determined by the sensitivity of our apparatus. However, there is a
 hint of increased surface potential fluctuation at frequencies  below
 $0.1\;\mHz$. This low frequency rise is found  consistently in similar
 measurements.

\section{Conclusion}

We presented the specifications and capabilities of our high
sensitivity torsion balance for LISA. In the absence of external
forces (i.e. large separations) the torsion balance reaches the
thermal limit given by the torsion fiber properties. 
At 1~mHz the LISA-equivalent acceleration
sensitivity is only one order of magnitude higher than the LISA
requirement. In addition, we have demonstrated that we can collect
data in an electrostatic feedback mode, where the pendulum is held at
a fixed angular position without increasing the noise.
We have measured the surface potential 
using this feedback method. We were able to measure the fluctuation of the
surface potential at sub-mHz frequencies. Currently, we are focusing
our efforts on improving the sensitivity for time varying surface potentials.

This work was supported by the NASA contracts NAS5-03075, NNC04GBG03G
and grant NNG05GF74G.

\bibliographystyle{aipproc}
\bibliography{lisa_sts}

\begin{thebibliography}{5}
\expandafter\ifx\csname natexlab\endcsname\relax\def\natexlab#1{#1}\fi
\providecommand{\enquote}[1]{``#1''}
\expandafter\ifx\csname url\endcsname\relax
  \def\url#1{\texttt{#1}}\fi
\expandafter\ifx\csname urlprefix\endcsname\relax\def\urlprefix{URL }\fi
\providecommand{\eprint}[2][]{\url{#2}}

\bibitem[Bender et~al.(1998)]{PPA}
P.~L. Bender, K.~V. Danzmann, and the {LISA} Study~Team, \emph{Laser
  Inteferometer Space Antenna for the Detection of Gravitational Waves,
  Pre-Phase A Report}, Max-Planck Institute for Quantum Optics, Garching,
  Germany, 1998, {MPQ}-233 2nd edn.

\bibitem[{DRS ITAT}(2005)]{DRS_ITAT}
{DRS ITAT}, \emph{{LISA} {DRS} Acceleration Noise Budget}, LIST Working Group
  3, 2005, {January} 25.

\bibitem[Hammesfahr(2000)]{FTR}
A.~Hammesfahr, \emph{{LISA}: Study of the Laser Interferometer Space Antenna,
  {\it {Final} {Technical} {Report}}}, ESTEC contract 13631/99/NL/MS, Astrium,
  2000.

\bibitem[Pollack et~al.(2006)]{Scott06}
S.~E. Pollack, S.~Schlamminger, and J.~H. Gundlach, \enquote{Outgassing,
  Temperature Gradients and the Radiometer Effect in {LISA}: A Torsion Pendulum
  Investigation,} in \emph{Proceedings of the 6th International {LISA}
  Symposium}, American Institute of Physics, 2006.

\bibitem[{Saulson}(1990)]{Saulson90}
P.~R. {Saulson}, \emph{Phys. Rev. D} \textbf{42}, 2437--2445 (1990).

\end{thebibliography}

\end{document}